\def\be{\begin{equation}}
\def\ee{\end{equation}}
\def\ba{\begin{eqnarray}}
\def\ea{\end{eqnarray}}
\def\l{\label}

\documentclass[12pt]{article}

\usepackage{epsfig}

\begin{document} 
 \title{Virtual photon impact factors with exact gluon kinematics} 

\author{A.Bialas \\ M.Smoluchowski Institute of Physics \\Jagellonian
University, Cracow\thanks{Address: Reymonta 4, 30-059 Krakow, Poland;
e-mail:bialas@th.if.uj.edu.pl}\\Institute of Nuclear Physics, Cracow \\
\\
H.Navelet and R. Peschanski \\ Service Physique Theorique,
 CE Saclay\thanks{Address: SPhT  CEA-Saclay  F-91191
Gif-sur-Yvette Cedex, France;  \newline emails: navelet@spht.saclay.cea.fr, 
pesch@spht.saclay.cea.fr}}
\maketitle

\begin{abstract}

An explicit analytic formula for the transverse and longitudinal impact factors 
$S_{T,L}(N,\gamma)$ of the photon using $k_T$
factorization with exact gluon kinematics is given. Applications to the QCD 
dipole model and the extraction of the unintegrated gluon structure function 
from data are proposed. 

\end{abstract}

\section{ Introduction}
In the 
present knowledge on perturbative QCD resummations  at leading level in 
logarithms of the energy (i.e. 
$log 1/x_{Bj}$) the coupling of external 
sources, in particular a virtual 
gluon in deep-inelastic reactions, is  based on the theorem of $k_T$ 
factorization \cite{cia}, proven in the 
leading
logarithmic approximation of QCD \cite{cia}. The theorem states that the
``unintegrated'' gluon distribution, i.e. the distribution of energy {\it and}
transverse momentum of gluons in the target, factorizes from the rest of
the process. The remaining factor is the so-called ``impact factor''. 
Consequently, this  ``impact factor'' is an
universal quantity, the same in all processes initiated by the same external 
source, e.g. the photon. 
The ``unintegrated'' gluon distribution will depend on the target, but again the 
target impact factor 
can also be factorized out, leaving place to an universal interaction term, 
given by the Balitsky, Fadin, Kuraev Lipatov BFKL Pomeron   \cite{bfkl}. At 
next-to-leading level, the modified 
interaction term is now known \cite{next} but not yet are the impact factors 
determined at this order of perturbation \cite{bar}. It is expected that the 
effect of the exact kinematics of the exchanged gluons, which is our subject,  
gives the main contribution to these higher order terms.

The $k_T$ factorized impact factors can be conveniently expressed in terms of 
two Mellin variables, $\gamma,$ conjugated to transverse momentum squared and 
$N,$ conjugated to energy. Up to now, only the impact factors at $N=0$ were 
considered, since, strictly speaking, $k_T$ factorization has been proven only 
at infinite energy, implying $N=0.$ If, however the validity of $k_T$ 
factorization is extended to the case of exact gluon kinematics, it implies the 
knowledge of the combined $\gamma,N$ dependence of impact factors. To our 
knowledge, this combined dependence has been derived only for real 
photoproduction of heavy flavors \cite{cat}. It is the purpose of the present 
paper to give an explicit analytic expression for the 
$\gamma,N$ dependence of virtual photon impact factors. To this
end, using the $k_T$ factorization and exact gluon kinematics, we derive
the explicit formulae for the total cross section of longitudinal and
transverse photons on any target with a given distribution of gluons.

In the next section 2, the formulae for the (virtual) photon impact factors  
following from $k_T$ factorization are given. The details of the calculation, 
implying multidimensional integration and resummation of generalized 
hypergeometric functions are presented in section 3 for the longitudinal case 
and in section 4 for the transverse one. Applications of our results are given 
in section 5.1 for the QCD dipole model \cite{fra} which turned 
out to be rather
successful in description of the deep inelastic total and diffractive
cross-section of the virtual photons \cite{npr}. In subsection 5.2 we 
suggest a model independent method of  extraction from data of the unintegrated 
gluon structure function. Our conclusions are given
in the last section.

\section{An explicit formula for the total cross -sections using  $k_T$
factorization}

Our  starting point   is the formula for the total longitudinal and 
transverse photon cross-sections given in \cite{kms}:
\ba
\sigma_L\equiv\frac{4\pi^2\alpha}{Q^2} F_L=4 \alpha \alpha_s Q^2
\int_0^1dz [z(1-z)]^2\nonumber \\
\int \frac{dk^2}{k^4} \int d^2p \left(\frac{1}{p^2+\hat{Q}^2}
-\frac{1}{(p-k)^2+\hat{Q}^2}\right)^2 g(x_g,k^2)\ .
\l{x1}
\ea
Using
\be
g(x_g,k^2)=\int\frac{dN}{2\pi i}(x_g)^{-N} \int \frac{d\gamma}{2\pi i}
\left( k^2\right)^{\gamma}\ \tilde{g}_N(\gamma)\ ,  \l{x2}
\ee
we write
\ba
\sigma_L=\int\frac{dN}{2\pi i}(x_{Bj})^{-N} 
\int \frac{d\gamma}{2\pi i}\ \tilde{g}_N(\gamma)
S_L(N,\gamma)\ (Q^2)^{\gamma-1}\l{ax2}
\ea
with
\ba
S_L(N,\gamma)=4\alpha \alpha_s 
\int_0^1dz [z(1-z)]^2
\int {dk^2}\ \left(\frac{k^2}{Q^2}\right)^{\gamma-2} \nonumber \\ \int d^2p 
\left(\frac{1}{p^2+\hat{Q}^2}
-\frac{1}{(\vec{p}-\vec{k})^2+\hat{Q}^2}\right)^2 
\frac{(\hat{Q}^2)^N}{[(\vec{p}-(1-z)\vec{k})^2+\hat{Q}^2+\hat{k}^2]^N}
\l{x3}
\ea
where we have used the relation
\be
x_g= x_{Bj}
\frac{(\vec{p}-(1-z)\vec{k})^2+\hat{Q}^2+\hat{k}^2]}{\hat{Q}^2}
;\;\;\;\;\;\hat{Q}^2=z(1-z)Q^2         \l{xx4}
\ee
Similarly, 
\ba
\sigma_T=\frac{4\pi^2\alpha}{Q^2} F_T= \alpha \alpha_s
\int_0^1dz [z^2+(1-z)^2]\nonumber \\
\int \frac{dk^2}{k^4} \int d^2p \left(\frac{\vec{p}}{p^2+\hat{Q}^2}
-\frac{\vec{p}-\vec{k}}{(\vec{p}-\vec{k})^2+\hat{Q}^2}\right)^2 g(x_g,k^2)
\l{1}
\ea
and using (\ref{x2}) and (\ref{xx4}) we have
\ba
\sigma_T\equiv\int\frac{dN}{2\pi i}(x_{Bj})^{-N} 
\int \frac{d\gamma}{2\pi i}\ \tilde{g}_N(\gamma)  S_T(N,\gamma)\ 
(Q^2)^{\gamma-1}
\l{d1}
\ea
with
\ba
S_T(N,\gamma)=\alpha \alpha_s
\int_0^1dz [z^2+(1-z)^2]
\int \frac{dk^2}{k^2}\
 \left(\frac{k^2}{Q^2}\right)^{\gamma-1} \nonumber \\ \int d^2p 
\left(\frac{\vec{p}}{p^2+\hat{Q}^2}
-\frac{\vec{p}-\vec{k}}{(\vec{p}-\vec{k})^2+\hat{Q}^2}\right)^2
\frac{(\hat{Q}^2)^N}{[(\vec{p}-(1-z)\vec{k})^2+\hat{Q}^2+\hat{k}^2]^N}\ .
  \l{3}
\ea
It turns out that the integrals (\ref{x3}) and (\ref{3}) can be
explicitly performed and expressed in terms of $\psi$ digamma functions.
 The details of the calculation are given in sections 3 and 4.
Here we only quote the final formulae:

\ba 
S_L(N,\gamma)=8\alpha \alpha_s 
\frac{\pi\Gamma(\gamma\!+\!\delta+1)\Gamma(\gamma\!+\!1)}{\Gamma(N)}\ 
\nonumber\\
 \frac{1}{(\delta^2\!-\!1)(\delta^2\!-\!4)}
\left\{ 
\frac{\psi(\gamma\!+\!\delta)-\psi(\gamma)}{\delta} \times 
\frac{3N^2\!-\!(\delta^2\!-\!1)} 
{2N} - 3 \right\}\  
\l{yiy}
\ea

\ba
S_T(N,\gamma)=2\alpha \alpha_s \frac 
{\pi\Gamma(\gamma\!+\!\delta)\ \Gamma(\gamma)}{\Gamma(N)} 
\frac 1{(\delta^2-1)(\delta^2-4)}\nonumber\\ \left\{\frac 
{\psi(\gamma\!+\!\delta)-\psi(\gamma)}{\delta} \times
\frac{N^2(3(N\!+\!1)^2\!+\!9)\!-\!2N(\delta^2\!-\!1)+(\delta^2\!-\!1)(\delta^2\!
-\!9)} 
{4N} 
\right.\nonumber\\
\left.-\frac12\left (3(N\!+\!1)^2\!+\!9\!+\!(\delta^2\!-\!1)\right)\right\} \ 
, 
\l{yjy}
\ea
where we adopted the convenient notation $\delta \equiv N-2\gamma+1,$

Note that the poles at $\delta=0,\pm1,\pm2$ in formulae 
(\ref{yiy},\ref{yjy}) are actually absent due to zeroes in the numerators, as it 
should from regularity of the generalized (Meijer) hypergeometric functions 
appearing in the derivation, see later. This provides numerous 
and non trivial checks of the resummations leading to (\ref{yiy},\ref{yjy}).

\section{Longitudinal photon impact factor}

As the first step in the calculation
 we observe that, using the symmetry of the integrand with 
respect to interchange of $z$ and $1-z$, the integrals in (\ref{x3}) can be
written as a sum of two terms:
\be
S_L(N,\gamma)=8\alpha \alpha_s \left(Q^2\right)^{2-\gamma} (A-B) \l{s1}
\ee
where
\ba 
A=
\int_0^1dz [z(1-z)]^2
\int \frac{dk^2}{k^4}\left(k^2\right)^{\gamma}\nonumber \\ \int d^2p 
\frac{1}{(p^2+\hat{Q}^2)^2}
\frac{(\hat{Q}^2)^N}{[(\vec{p}-(1-z)\vec{k})^2+\hat{Q}^2+\hat{k}^2]^N}
\l{ss3}
\ea
\ba 
B=
\int_0^1dz [z(1-z)]^2
\int \frac{dk^2}{k^4}\left(k^2\right)^{\gamma} \nonumber \\ \int d^2p 
\frac{1}{[p^2+\hat{Q}^2][(p-k)^2+\hat{Q}^2]} 
\frac{(\hat{Q}^2)^N}{[(\vec{p}-(1-z)\vec{k})^2+\hat{Q}^2+\hat{k}^2]^N}\ .
\l{sss3}
\ea
Using several times the identity
\be
\frac1{C^M} = \frac1{\Gamma(M)}\int_0^{\infty}dt \ t^{M-1} e^{-tC}  \l{s41}
\ee
we transform (\ref{ss3}) and (\ref{sss3}) into
\ba 
A= \frac1{\Gamma(N)}
\int_0^1dz [z(1-z)]^2(\hat{Q}^2)^N
\int \frac{dk^2}{k^4}\left(k^2\right)^{\gamma} 
\int vdv \int dt\ t^{N-1}\nonumber \\ \int d^2p\ 
\exp\{-v(p^2\!+\!\hat{Q}^2)\!-\!t
[(\vec{p}\!-\!(1\!-\!z)\vec{k})^2\!+\!\hat{Q}^2\!+\!\hat{k}^2]\} \ ,\l{4c}
\l{ss4}
\ea
\ba 
B=\frac1{\Gamma(N)}
\int_0^1dz [z(1-z)]^2 (\hat{Q}^2)^N
\int \frac{dk^2}{k^4}\left(k^2\right)^{\gamma} 
\frac1{\Gamma(N)}\nonumber \\\int dv \int dv' \int dt\ t^{N\!-\!1} 
\int d^2p\  
e^{\!-\!v(p^2\!+\!\hat{Q}^2)\!-\!v'((p\!-\!k)^2\!+\!\hat{Q}^2)
\!-\!t[(\vec{p}\!-\!(1\!-\!z)\vec{k})^2\!+\!\hat{Q}^2\!+\!\hat{k}^2]} 
\l{s42}
\ea
Thus the integration over $d^2p$ reduces to a gaussian form and can be 
easily performed. After rescaling the variables $v=ty,v'=ty'$
and substitution $u=tk^2$ the integrals over $du$ and $dt$ factorize
from the rest and can be explicitly evaluated. The final result of these
operations reads
\ba
A= 
\frac{\pi\Gamma(\gamma-1)\Gamma(N-\gamma+2)}{\Gamma(N)}(Q^2)^{\gamma-2}
\int_0^1dz z^{\gamma}(1-z)\nonumber \\
\int \frac{ydy}{(1+y)^{N-2\gamma+4}}(y+z)^{1-\gamma}\ ,
 \l{x7}
\ea
\ba
B= 
\frac{\pi \Gamma(N-\gamma+2)\Gamma(\gamma-1)}{\Gamma(N)}(Q^2)^{\gamma-2}
\int_0^1dz  [z(1-z)]^{\gamma} \nonumber \\
\int \frac{dy dy'[(y+z)(y'+1-z)]^{1-\gamma}}{(1+y+y')^{N-2\gamma+4}}\ .
 \l{x14}
\ea
To evaluate integrals in (\ref{x7}) one takes $h=1\!+\!y,\ y\!+\!z = 
h[1\!-\!(1\!-\!z)/h]$. Using the formula for the Gauss series
\ba
(1-x)^{-a} =\sum_{n=0}^{\infty} \frac{\Gamma(a+n)}{\Gamma(a)}\frac
{x^n}{n!} \  \l{s5}
\ea
and picking the factor $\Gamma(\gamma-1)$ in formula (\ref{x14}), the integral 
over 
$dv$ leads to 
\ba
\Gamma(\gamma\!-\!1)&&\!\!\!\!\int 
\frac{ydy}{(1\!+\!y)^{N\!-\!2\gamma\!+4\!}}(y\!+\!z)^{1\!-\!\gamma}
=\nonumber \\ 
&=&\sum\frac{\Gamma(\gamma\!-\!1\!+\!n)}{(N\!-\!\gamma\!+\!2\!+\!n)(N\!-\!\gamma
\!
+\!1\!+\!n)}\frac{(1\!-\!z)^n}{n!}
\nonumber \\    &\equiv& \  _2G_1 
(\gamma\!-\!1,N\!-\!\gamma\!+\!1;N\!-\!\gamma\!+\!3;1\!-\!z)\ ,
  \l{x8}
\ea
where the Meijer function $_2G_1(u,v;w;t)\equiv  
{\Gamma(u)\Gamma(v)}/{\Gamma(w)}\ _2F_1(u,v;w;t).$

Substituting this into (\ref{x7}) we can integrate over $z$ to obtain
\be
A=
\frac{\pi\Gamma(N-\gamma+2)\Gamma(\gamma+1)}{\Gamma(N)}(Q^2)^{\gamma-2}\ T_A
\l{xx10}
\ee
where
\ba
T_A=
 \ _3G_2 
(2,\gamma-1,N-\gamma+1;\gamma+3,N-\gamma+3;1)\ .
\l{x10}
\ea
To calculate the integrals over $dy$ and $dy'$ in (\ref{x14}) we first
rescale the variables $y,y'$ $\rightarrow$ $zy,(1-z)y'$. Then we
introduce $h=1+y,h'=1+y'$ and, finally $\xi=h/h'$. The last change
implies $1+y+y'\rightarrow 1+zy+(1-z)y'=h'[1-z(1-\xi)]$.
 Using again (\ref{s5}) the integral
over $dy dy'$ can be transformed into the series
\ba
\int \frac{dy dy'[(y+z)(y'+1-z)]^{1-\gamma}}
{(1\!+\!y\!+\!y')^{N\!-\!2\gamma\!+\!4}}= \nonumber \\
=[z(1-z)]^{2-\gamma}
\frac{\Gamma(N\!+\!2\!-\!\gamma)}{N\Gamma(N\!-2\gamma\!+\!4)}\sum_{n=0}^{\infty}
\frac{\Gamma(N\!-\!2\gamma\!+\!4\!+\!n)}{\Gamma(N\!-\!\gamma\!+\!3\!+\!n)} 
\left(z^n \!+\!(1\!-\!z)^n\right) 
\nonumber \\
\equiv [z(1\!-\!z)]^{2\!-\!\gamma}
\frac{\Gamma(N\!+\!2\!-\!\gamma)}{N\Gamma(N\!-\!2\gamma\!+\!4)}\ 
\left[_2G_1(1,N\!-\!2\gamma\!+\!4;N\!-\!\gamma\!+\!3;z)\right.\nonumber \\
\left.+(z\rightarrow 1\!-\!z)\right]\ .
\l{x15}
\ea
Using known relations \cite{grad} on $_2F_1$ functions, one  writes also
\ba
\int \frac{dy dy'[(y+z)(y'+1-z)]^{1-\gamma}}
{(1+y+y')^{N-2\gamma+4}}= \nonumber \\
=z^{2-\gamma}
\frac1{N\Gamma(\gamma-1)}\ 
\left\{_2G_1(\gamma\!-\!1,N\!-\!\gamma\!+\!2;N\!-\!\gamma\!+\!3;z)+(z\rightarrow
1\!-\!
z)\right\}\ .
\l{x155}
\ea

Introducing this into (\ref{x14}) we  perform integration over $z$
to obtain finally
\ba
B\equiv \frac{2\pi \Gamma(N-\gamma+2)\Gamma(\gamma+1)}{\Gamma(N)N}
(Q^2)^{\gamma-2}\ T_B \l{x17}
\ea
where
\be
T_B=
(Q^2)^{\gamma-2} \ _3G_2(3,\gamma-1,N-\gamma+2;\gamma+4,N-\gamma+3;1)\ . 
\l{x175}
\ee

It turns out that the series in (\ref{x10}) and(\ref{x175}) can be summed
up and expressed in terms of the digamma functions.  One uses known relations on 
Meijer 
functions at $z=1,$ see 
ref.\cite{prud}. One writes 
\ba
T_A   \equiv \ _3G_2 (2,a,b;a+\!4,b\!+\!2;1)=\nonumber\\
\ _3G_2 (2,a,b;a\!+\!4,b\!+\!1;1)- \ _3G_2(2,a,b\!+\!1;a\!+\!4,b\!+\!2;1) 
=\nonumber\\
 \frac13 \left[(b\!-\!1)\ _3G_2 (1,a,b\!-\!1;a\!+\!3,b\!+\!1;1)- b\ 
_3G_2(1,a,b;a\!+3,b\!+\!2;1)\right]\ .
\l{x176}
\ea
and 
\ba
T_B   \!\equiv \!\!\ _3G_2 
(3,a,b\!+\!1;a\!+\!5,b\!+\!2;1)\!=\!\frac{b(b\!-\!1)}{6} \ _3G_2 
(1,a,b\!-\!1;a\!+\!3,b\!+\!2;1)\ ,
\l{x177}
\ea
using the abbreviations
\be
a=\gamma-1;\;\;\;\; b=N+1-\gamma\ .\;\;\;\;
\l{4}
\ee

Using the generic formula for hypergeometric functions $_3F_2 
(1,a,b;a\!+\!p\!+\!1,b\!+\!q\!+\!1;1)$ \cite{muna}, one writes:
\ba
_3G_2 (1,a,b;a\!+\!p\!+\!1,b\!+\!q\!+\!1;1) \equiv \nonumber\\(-1)^{p\!+\!1} 
\frac 
{\Gamma(p\!+\!q\!+\!1)}{\Gamma(p\!+\!1)\Gamma(q\!+\!1)}\frac{\Gamma(b\!-\!a\!-\!
p)}{\Gamma(b\!-\!a 
\!+\!q\!+\!1)}
\left(\psi(a)-\psi(b)\right) + \nonumber\\
\left[(-1)^{p}\frac 
{\Gamma(b\!-\!a\!-\!p)}{\Gamma(1\!-\!a)\Gamma(q\!+\!1)} 
 \sum_{k=0}^{p\!-\!1}\frac1{k\!-\!p}\frac1{\Gamma(k\!+\!1)}\frac 
{\Gamma(1\!+\!q\!+\!k)\Gamma(1\!-\!a\!-\!p\!+\!k)}{\Gamma(b\!-\!a\!+\!q\!-\!p
\!+\!k\!+\!1)} \right.\nonumber\\ \left.+
\left\{a<\!-\!>b,p<\!-\!>q\right\}\right]\ .
\l{x178}
\ea
All in all, using for convenience 	 the notations $\delta \equiv 
b-a-1$ and $N\equiv b+a,$ we get

\ba  
 T_A-\frac{2T_B}{N}=\frac1{(\delta^2\!-\!1)(\delta^2\!-\!4)}\left( 
\frac{\psi(a\!+\!1\!+\!\delta)\!-\!\psi(a\!+\!1)}{\delta}\ 
\frac{3N^2\!-\!(\delta^2\!-\!1)} 
{2N} - 3\right) .
 \l{finL}
\ea

\section{Transverse photon impact factor}
$\sigma_T$ can be evaluated  along the similar lines as $\sigma_L$. 
Here we mark only the main differences. The key point is that  $\sigma_T$ can be 
evaluated using $\sigma_L,$ which 
gives a noticeable simplification of the  painful calculation.

Let us come back to the calculation of 
$\sigma_L,$ starting with formula (\ref{s1}) and define 
\be
(DL) \equiv Q^2 A\ ; \ (NDL)  \equiv Q^2 B\ .
\l{def}
\ee
Using (\ref{3}) we write
\be
S_T(\gamma,N)= 2\alpha \alpha_s ((DT)-(NDT)) \left(Q^2\right)^{1-\gamma}  \l{m1}
\ee
with
\ba
(DT)=\int_0^1dz [z^2+(1-z)^2]\int
dk^2(k^2)^{\gamma-2}  \nonumber \\
\int d^2p \frac{(\vec{p})^2}{(p^2+\hat{Q}^2)^2}
\frac{(\hat{Q}^2)^N}{[(\vec{p}-(1-z)\vec{k})^2+\hat{Q}^2+\hat{k}^2]^N}
 \l{a3a}
\ea  
\ba
(NDT)=\int_0^1 dz [z^2+(1-z)^2] 
\int dk^2(k^2)^{\gamma-2}  \nonumber \\
 \int d^2p \frac{\vec{p}(\vec{p}-\vec{k})}{[p^2+\hat{Q}^2]
[(p-k)^2+\hat{Q}^2]}
\frac{(\hat{Q}^2)^N}{[(\vec{p}-(1-z)\vec{k})^2+\hat{Q}^2+\hat{k}^2]^N}\ .
\l{m3}
\ea
Let us rewrite the quantities of interest in the following form:
\ba
(DL) &\equiv & \int_0^1 dz A_L(z) \ I(z)\nonumber \\
(NDL) &\equiv & \int_0^1 dz A_L(z) \ J(z)\nonumber \\
(DT) &\equiv & \int_0^1 dz A_T(z) \ [K(z) - I(z)]\nonumber \\
(NDT) &\equiv & \int_0^1 dz A_T(z) \ [K(z) - J(z)-\frac 12 L(z)]\ ,
\l{defintegrales}
\ea
where
\be
A_L(z)=z(1-z)\ ; A_T(z)=z^2 + (1-z)^2=1-2A_L(z)\ . 
\l{A}
 \ee
The integrals $I,J,K,L$ are defined as follows:
\ba
I(z) = \int \frac{dk^2}{k^4}k^{2\gamma} \int d^2p 
\frac{\hat{Q}^2}{(p^2+\hat{Q}^2)^2}
\frac{(\hat{Q}^2)^N}{[(\vec{p}-(1-z)\vec{k})^2+\hat{Q}^2+\hat{k}^2]^N}\ ,
\l{I}
\ea 
\ba
J(z) = \int \frac{dk^2}{k^4}k^{2\gamma}  \int d^2p 
\frac{\hat{Q}^2}{[p^2\!+\!\hat{Q}^2][(p\!-\!k)^2\!+\!\hat{Q}^2]}
\frac{(\hat{Q}^2)^N}{[(\vec{p}\!-\!(1\!-\!z)\vec{k})^2\!+\!\hat{Q}^2\!+\!\hat{k}
^2]^N}\ ,
\l{J}
\ea
\ba
K(z) = \int \frac{dk^2}{k^4}k^{2\gamma} \int d^2p 
\frac{1}{(p^2+\hat{Q}^2)}
\frac{(\hat{Q}^2)^N}{[(\vec{p}-(1-z)\vec{k})^2+\hat{Q}^2+\hat{k}^2]^N}\ ,
\l{K}
\ea 
\ba
L(z) \!=\! \int \!\frac{dk^2}{k^4}k^{2\gamma} \!\int d^2p 
\frac{k^2}{[p^2+\!\hat{Q}^2][(p\!-\!k)^2\!+\!\hat{Q}^2]}
\frac{(\hat{Q}^2)^N}{[(\vec{p}\!-\!(1\!-\!z)\vec{k})^2\!+\!\hat{Q}^2\!+\!\hat{k}
^2]^N}\ .
\l{L}
\ea
Consequently, one reads for  $\sigma_L:$
\be
(DL)-(NDL)=\int_0^1 A_L(z) 
[I(z)-J(z)]
 \l{s2}\ .
 \ee
 The corresponding expression for $\sigma_T$ shows a nice simplification, since 
the integrand $K(z)$ cancels from the calculation:
 \be
(DT)-(NDT)=\int_0^1 A_T(z) 
[J(z)-I(z)+\frac12 L(z)]
 \l{s3}\ .
 \ee

 Using (\ref{A}), we get
 \ba
 (DT)-(NDT) =  2[(DL)-(NDL)] + \nonumber\\
 + \int_0^1dz 
[J(z)-I(z)] +\frac 12\int_0^1dz (1-2z(1-z)) L(z)\ .
 \l{s4}
 \ea
 From formula (\ref{x8}), it is straightforward to obtain (in the $a,b$ 
notations):
 \be
 I(z)=Q^{2a}\frac{\Gamma(b+1)}{\Gamma(b+a)}z^a \ _2G_1(a,b;b+2;1-z)\ ,\l{g1}
 \ee
 where $_2G_1(a,b;b+2;1-z)\equiv {\Gamma(a)\Gamma(b)}/{\Gamma(b+2)}\  
_2F_1(a,b;b+2;1-z).$
 
 After some  straightforward transformations from formula 
(\ref{x17}), one writes 
  \be
 J(z)=Q^{2a}\frac{\Gamma(b\!+\!1)}{\Gamma(b\!+\!a\!+\!1)}z(1\!-\!z)[z^{a\!-\!1} 
\ _2G_1(a,b\!+\!1;b\!+\!2;1\!-\!z)+(z\rightarrow1\!-\!z)]\ ,\l{g2}
 \ee
 \be
 L(z)=Q^{2a}\frac{\Gamma(b)}{\Gamma(b+a+1)}[z^{a} 
\ _2G_1(a+1,b;b+1;1-z)+(z\rightarrow1-z)]\ ,\l{g3}
 \ee 
 where $L(z)$ is obtained from $J(z)$ by  suppressing the factor $z(1-z)$ and 
replacing $a\to a+1, b\to b-1,$  due to the numerator $k^2$ instead of 
$\hat{Q}^2$ for $J(z)$ in formula 
(\ref{L}).

 Inserting expressions (\ref{g1},\ref{g2},\ref{g3}) in formula (\ref{s4}), 
one gets after integration over $z$:
\ba
 (DT)-(NDT)= 2 [(DL)-(NDL)] + \frac {\Gamma(b)\Gamma(a+1)}{\Gamma(a+b+1)} Q^{2a} 
\nonumber\\
\times\ \left[_3G_2(1,a+1,b;a+2,b+1;1)-ba\ _3G_2(1,a,b;a+2,b+2;1)\right. 
\nonumber\\
\left.-(b-1)(a+1)_3G_2(1,a+1,b-1;a+3,b+1;1)\right]\ .
\l{final}
\ea
All in all, using for convenience 	 the notations $\delta \equiv b-a$ and 
$N\equiv b+a,$
one finally obtains
\ba 
(DT)-(NDT)=\frac {\Gamma(b)\Gamma(a\!+\!1)}{\Gamma(a\!+\!b)} Q^{2a}
\frac 1{(\delta^2\!-\!1)(\delta^2\!-\!4)}    \nonumber\\
\left\{\frac{\psi(a\!+\!1\!+\!\delta)\!-\!\psi(a\!+\!1)}{\delta} \right.\! 
\times \! 
 \frac{N^2(3(N\!+\!1)^2\!+\!9)\!-\!2N(\delta^2\!-\!1)\!+
\!(\delta^2\!-\!1)(\delta
^2\!-\!9)} {4N} 
 \nonumber\\ -\left.\!\frac12\left( 
3(N\!+\!1)^2\!+\!9\!+\!(\delta^2\!-\!1)\right)\right\}
\l{finT}
\ea
\section{Applications}

\subsection{Comparison with the dipole model}

In a recent paper \cite{bnp0} we have shown that  the dipole model is, 
strictly speaking, not
compatible with the formulae obtained from the $k_T$ factorization
including the exact kinematics of the corresponding Feynman diagrams.
The point is that the results from $k_T$ factorization are non-diagonal
in impact parameter space, contrary to the fundamental assumption of the
dipole model. Thus, it is worthwhile to derive the explicit modifications of the 
model due to the gluon kinematics.

The formulae of the previous section can now be compared with those
obtained in the QCD dipole model, which amounts to consider \cite{npr} the 
impact factors at $N=0.$:

\ba
S_L^{dip}(\gamma)\equiv S_L(N\!=\!0,\gamma)=\frac{\pi^2\alpha \alpha_s 
}{3}\frac{\gamma(1\!-\!\gamma)}
{1\!-\!\frac23\gamma}\frac{\Gamma^2(1\!-\!\gamma)\Gamma^2(\gamma)}
{\Gamma(\frac32\!-\!\gamma)\Gamma(\frac32\!+\!\gamma)}
\l{ada}
\ea
\ba
S_T^{dip}(\gamma)\equiv S_T(N\!=\!0,\gamma)=\frac{\pi^2\alpha \alpha_s 
}{3}\frac{(1\!+\!\gamma)(1\!-\!\frac12\gamma)}
{1\!-\!\frac23\gamma}\frac{\Gamma^2(1\!-\!\gamma)\Gamma^2(\gamma)}
{\Gamma(\frac32\!-\!\gamma)\Gamma(\frac32\!+\!\gamma)}\ .
\l{bdb}
\ea

The knowledge of $S_{L,T}(N,\gamma)$ allows one to take into account  the 
modifications of the QCD dipole model due to exact gluon kinematics. 
Indeed, let us insert the BFKL pole in the formula for the unintegrated gluon 
structure 
function 
\be
\tilde{g}_N(\gamma) =\frac 
{\bar{v}(\gamma)w(\gamma)}{N\!-\!\bar\alpha\chi(\gamma)} 
\left(Q_0^2\right)^{-\gamma} \ ;\chi(\gamma)\equiv 
2\psi(1)-\psi(\gamma)-\psi(1\!-\!\gamma)\ ,
\label{bfkl}
\ee
where $\bar\alpha$ is the effective value of the strong coupling constant 
$\bar\alpha=\frac {\alpha_s N_c}{ \pi},$ in the BFKL
kernel, $w(\gamma)$ is the Mellin transform of 
the probability to find a dipole in the target,  $Q_0$ sets the typical 
model scale of the unintegrated gluon structure function
$g(x_g,k^2),$ (see formula (\ref{x2})) and 
\be
\bar{v}(\gamma) =  \frac{\bar{\alpha}}{
2^{2\gamma}\gamma}\frac{\Gamma(1-\gamma)}{\Gamma(1+\gamma)}  \l{bfkl1}
\ee
is the Mellin transform of the probability 
to find a gluon in a dipole.
In its final form, the modified QCD dipole model for longitudinal (\ref{ax2}) 
and transverse 
(\ref{d1}) cross sections with full kinematics read:
\be
\sigma_L=Q^{-2}\int \frac{d\gamma}{2\pi i}\ (x_{Bj})^{-\bar\alpha\chi(\gamma)} 
\bar{v}(\gamma)w(\gamma)
S_L(N\!=\bar\alpha\chi(\gamma),\gamma)\ \left(\frac 
{Q^2}{Q_0^2}\right)^{\gamma}\ , 
\l{adaf}
\ee 
\be
\sigma_T=Q^{-2}
\int \frac{d\gamma}{2\pi i} (x_{Bj})^{-\bar\alpha\chi(\gamma)}   
\bar{v}(\gamma)w(\gamma)
S_T(N\!=\bar\alpha\chi(\gamma),\gamma)\ \left(\frac 
{Q^2}{Q_0^2}\right)^{\gamma}
\ .
\l{d1f}
\ee
Note that in these formulae, one has $\delta \equiv N-2\gamma+1= 
\bar\alpha\chi(\gamma)-2\gamma+1.$

Formulae (\ref{adaf}) and (\ref{d1f}) give the explicit dependence of the 
cross-sections in the shift of the hard pomeron intercept.

We have compared numerical results (in the saddle-point approximation)
from the Eqs (\ref{yiy}) and (\ref{yjy}) 
with those obtained from Eqs (\ref{ada}) and (\ref{bdb}) in the range
$.01>x>.0001$ and $20 Gev^2 <Q^2<160 Gev^2$. It turns out
that they give a rather similar dependence on both $x$ and $Q^2$ (the
deviations do not  exceed $5\%$). Normalization changes, however: the
cross-sections including the full gluon kinematics are by about factor
2 smaller than the ones obtained from the high-energy approximation. 
Note the interesting fact that the ratio $R=\sigma_L/\sigma_T$ is affected: it 
increases by about 
$15\%$.

It is not very surprizing that the dependence on kinematic variables
does not substantially differ in the two approaches. Indeed, this
dependence is mostly controlled by the position of the saddle point
which is the same in the two formulae. The rather important change in
normalization, however, could not have been easily guessed. In particular the 
ratio of normalizations in $R=\sigma_L/\sigma_T$ is a quantity of experimental 
interest.

\subsection{Method of extraction of the unintegrated gluon distribution}

The knowledge of the impact factors $S_L(N,\gamma)$ and $S_T(N,\gamma)$ as  
explicit analytic functions of their two variables allows a model independent 
determination of the unintegrated gluon structure function from $\sigma_L$ or 
$\sigma_T$ or from the  experimental observable $F_2\equiv 
\frac{Q^2}{4\pi^2\alpha}(\sigma_T+\sigma_L).$

Indeed,  formulae (\ref{ax2},\ref{d1}) yield
\ba
\tilde\sigma_{T,L}(N,\gamma)&\equiv& \int 
\frac{dx_{Bj}}{x_{Bj}}\left(x_{Bj}\right)^{N} 
\int dQ^2\left( {Q^2}\right)^{-\gamma}\sigma_{T,L}(x_{Bj},Q^2)\nonumber\\ &=& 
\tilde{g}_N(\gamma)  
S_{T,L}(N,\gamma)\ .
\l{d12}
\ea
Hence, from formula (\ref{x2}) defining the unintegrated gluon structure 
function, one gets 
\be
g(x_g,k^2)=\int\frac{dN}{2\pi i}(x_g)^{-N} \int \frac{d\gamma}{2\pi i}
\left( k^2\right)^{\gamma}\ \frac 
{\tilde\sigma_{T,L}(N,\gamma)}{S_{T,L}(N,\gamma)}\ .  
\l{x'2}
\ee
This relation (\ref{x'2}) has quite interesting features.

 First, it allows a 
determination of the unintegrated gluon structure function from experiments. For 
instance one may consider the ratio  
$(\tilde\sigma_T+\tilde\sigma_L)/(S_{T}+S_L)$ using data on $F_2.$

Second, the expected universality properties of $g(x_g,k^2)$ from $k_T$ 
factorization gives  predictions for  various processes e.g. the 
ratio $R=\frac {F_L}{F_T},$ the photoproduction and  leptoproduction of heavy 
flavors, diffractive leptoproduction of vector mesons and any other process 
where $k_T$ factorization applies. 

Third, the applicability of this relation 
goes 
beyond a specific model like the QCD dipole one, provided the coupling to the 
virtual photon comes from the exchange of two off-mass shell gluons with exact 
kinematics.

These results comes from the assumption that $k_T$ factorization is a valid 
approximation, e.g. the dominance of the two considered Feynman diagrams for the 
impact factors, when exact gluon kinematics can be considered. It remains to 
know to
what extent  higher order QCD contributions and non perturbative corrections may 
spoil  this approximation.

 \section{Conclusions}
 Our formulae (\ref{yiy},\ref{yjy}) give the exact two-variable dependence of 
the longitudinal and transverse impact factors of the photon in terms of the 
usual Mellin variables $N,\gamma$. $N$ is the variable conjugated to $x_{Bj},$ 
while $\gamma$ is conjugated to $Q^2.$ 
 In the particular framework of the QCD dipole model, 
it gives the modification taking into account the shift in $ 
N\!=\!\bar\alpha\chi(\gamma)$ of the BFKL singularity. This results  mostly in a 
change of the normalization of the cross-section, whereas the
dependence on kinematic variables predicted by the QCD dipole model is hardly 
affected. Note that the relative normalization in $R=\frac {F_L}{F_T}$ is 
increased by 15\%. 

 More generally, our result  opens the way  of a model-independent 
extraction of the universal unintegrated gluon 
 structure function which appears in various processes, whenever $k_T$ 
factorization can be applied.

 \vspace{0.3cm}
{\bf Acknowledgements}
\vspace{0.3cm}

A.B. thanks the Service de Physique Theorique of Saclay for support and
kind hospitality. This investigation was supported in part by the KBN
Grant No 2 P03B 086 14 and by the Subsidium of Fundation for Polish
Science 1/99.

\end{document}